# The impact of a large object with Jupiter in July 2009

Short Title: Impact on Jupiter in 2009

A. Sánchez-Lavega[1*], A. Wesley[2], G. Orton[3], R. Hueso[1], S. Perez-Hoyos[1], L. N. Fletcher[3], P. Yanamandra-Fisher[3], J. Legarreta[1], I. de Pater[4], H. Hammel[5], A. Simon-Miller[6], J. M. Gomez-Forrellad[7], J. L. Ortiz[8], E. García-Melendo[7], R. C. Puetter[9], P. Chodas[3]

* To whom correspondence should be addressed:
Dpto. Física Aplicada I, Escuela Superior de Ingenieros, Universidad del País Vasco, Alameda Urquijo s/n, 48013 Bilbao (Spain)
E-mail: agustin.sanchez@ehu.es

[1]Universidad del País Vasco, Bilbao (Spain)
[2]Acquerra Pty. Ltd. (Australia)
[3]Jet Propulsion Laboratory, Caltech (USA)
[4]University of California, Berkeley (USA)
[5]Space Science Institute (USA)
[6]NASA Goddard Space Flight Center (USA)
[7]Fundació Observatori Esteve Duran, Barcelona (Spain)
[8]Instituto de Astrofísica de Andalucía, C.S.I.C., Granada (Spain)
[9]Center for Astrophysics and Space Science, San Diego (USA).

## ABSTRACT

On 2009 July 19, we observed a single, large impact on Jupiter at a planetocentric latitude of 55°S. This and the Shoemaker-Levy 9 (SL9) impacts on Jupiter in 1994 are the only planetary-scale impacts ever observed. The 2009 impact had an entry trajectory opposite and with a lower incidence angle than that of SL9. Comparison of the initial aerosol cloud debris properties, spanning 4,800 km east-west and 2,500 km north-south, with those produced by the SL9 fragments, and dynamical calculations of pre-impact orbit, indicate that the impactor was most probably an icy body with a size of 0.5-1 km. The collision rate of events of this magnitude may be five to ten times more frequent than previously thought. The search for unpredicted impacts, such as the current one, could be best performed in 890-nm and K (2.03-2.36 μm) filters in strong gaseous absorption, where the high-altitude aerosols are more reflective than Jupiter's primary clouds.

Subject headings: planets and satellites: general, atmospheres, Jupiter

## 1. Introduction

Major impacts have modified the structure of Solar System bodies (de Pater & Lissauer, 2010) and changed the course of biological evolution on Earth (Kasting & Catling, 2003). With 70% the total mass of the planets, Jupiter is the major attractor for impacting bodies, and its massive atmosphere constitutes a natural laboratory for studying the impact response. In 1994 several fragments of comet Shoemaker-Levy 9 (SL9) impacted Jupiter between 16 and 22 July (Hammel et al. 1995; Harrington et al. 2004). The next such event was predicted to be hundreds of years in the future (Harrington et al. 2004). However, 15 years later a second large impact occurred. We have analyzed the impact debris in the discovery images to retrieve the impactor size, trajectory and impact time, constraining its pre-impact orbit and possible origin. We revise previous predictions on the impact rates with Jupiter and propose future search methods for their detection.

## 2. Debris Observations and Analysis

The dark impact- "bruise" was first noticed on CCD images of Jupiter obtained by Wesley on 2009 July 19 at 14:02 UT, just rotating into view from Jupiter's west limb (Figure 1). This feature, recorded by several amateur observers, was tracked during the next Jupiter rotation (July 20[th], ~ 01-02 UT) on images sent to the International Outer Planet Watch (IOPW) database (http://www.pvol.ehu.es/). The first images in methane and hydrogen absorption between 2.12- and 2.3- μm wavelength, were obtained at NASA's Infrared Telescope Facility (IRTF) during the third rotation after the impact (July 20[th], ~ 10-13 UT) (Figure 2a-b). They showed the spot to be very bright compared to the surroundings, indicating that the material high in the atmosphere at $P_{top}$ ~ 1-10 mbar (Hammel et al., 2010; Orton et al., 2010), i. e., above the main Jovian clouds ($P_{top}$ ~ 500 mbar). However, in visible light, the feature appeared dark against the main clouds. Because its visible and near-infrared morphology and reflectivity were so similar to the previous SL9 impact observations (Hammel et al. 1995; Harrington et al. 2004), the feature was most likely formed by the debris left by an impact.

A survey of amateur observations of Jupiter obtained between ~ 0.35 and 1 μm before the identification of the debris impact (see IOPW database) indicates that the spot was not present on July 19 as late as 7:40 UT, suggesting that the impact occurred between 7:40 and 14:02 UT. Similarity with young impact debris from SL9 (Figure 2) suggests the most probable impact time was 9 - 11 UT. The impact itself was not observed because it occurred on Jupiter's far side. IOPW images between June and September did not show any similar features at the same latitude bigger than ~ 200 km. We conclude that a single object impacted Jupiter on July 19, unlike the cometary fragments of SL9.

The center of the dark spot at continuum visible wavelengths was located at System-III longitude 304.5±0.5º and planetocentric latitude 55.1±0.5ºS, ~11-12º south of the SL9 impacts (Hammel et al. 1995). The initial feature consisted of two elements: A streak (the main spot) and a low-contrast extended crescent west of the main spot (Figure 2a), both dark in the visible and bright in the near infrared. This is similar to what was observed for SL9, but it is still a disputed issue how these features were generated by the impact (Crawford 1996; Mac Low 1996; Zahnle 1996; Takata & Ahrens, 1997; Harrington & Deming, 2001).

The streak had an elongated, approximately elliptical shape, with size along the major and minor axis of 6.7º in longitude (4,800 ± 200 km east-west) and 2º in latitude 2,500 ± 200 km north-south). This streak is tilted by 12 ± 2º in the northwest − southeast direction (Figure 2a) relative to the latitude circle passing through its geometric centre. This angle marks the approximate impactor entry direction with azimuth angle 290º (north is 0º, east is 90º, and so forth), as measured in orthographic projection (grid in km, Figure 2c). This is nearly opposite to the direction of the SL9 fragments, whose azimuths were all 164º.

The thin debris crescent northwest of the main spot extends 4800 km from the western edge of the streak, (8800 km from its centre). It is oriented with an azimuth 310º measured in the orthographic projection. Just as for the SL9 impacts (Pankine & Ingersoll 1999; Jessup et al. 2000), we interpret the 20º azimuthal clockwise rotation of the crescent, relative to the major axis of the streak, as due to the action of the Coriolis force on the falling material plus a sliding in the atmosphere that conserves the tangential velocity. To check this interpretation we present a simple model that constitutes a reasonable approach to the impact structure. At present, the available data, worse that those for the SL9, preclude a more sophisticated analysis.

The ballistic trajectories of the ejecta are given according to Jessup et al. (2000) by,

$$x(t) = \frac{1}{3}\omega\,g\,t^3\cos\lambda - \omega t^2\left(v_{0z}\cos\lambda - v_{0y}\sin\lambda\right) + v_{0x}t + x_0 \qquad (1)$$

$$y(t) = -\omega t^2 v_{0x}\sin\lambda + v_{0y}t + y_0 \qquad (2)$$

$$z(t) = -\frac{1}{2}g\,t^2 + \omega v_{0x}t^2\cos\lambda + v_{0z}t + z_0 \qquad (3)$$

Here $x_0$ and $y_0$ mark the impact location in Cartesian coordinates and $z_0$ represents the 100 mb altitude level, $x$ and $y$ are the coordinates taken east and north from the origin at time $t$ (see Figure 2c) and $z$ is the altitude above the $z_0 = 100$ mb level. The initial-velocity components in this reference frame are $\left(v_{0x}, v_{0y}, v_{0z}\right)$ and $\omega$ is Jupiter's angular rotation velocity. For simplicity we assumed a constant planetocentric latitude $\lambda = 55$ºS and constant Jovian gravitational acceleration $g = 25.902$ ms$^{-2}$. We computed ballistic trajectories that ascend and descend over a 1600-km horizontal distance, equal to the quasi-circular left boundary of the streak (Fig. 2c). Larger or smaller horizontal distances did not reproduce the final ejecta pattern. The horizontal distance provides a relation between the initial ejecta velocity $v_0$ and the elevation angle of the ejecta $\theta$ (measured from the vertical) given by $v_{0z} = v_0 \cos \theta$, with $v_{0x}$ and $v_{0y}$ also depending on the azimuth of the outgoing trajectory. In the ballistic trajectory the particles modify their velocity by the action of Coriolis forces. After falling back, we assume the ejecta bounces horizontally with only horizontal Coriolis forces. The equations of motions are modified to:

$$x(t) = x_f + v_{fx}t + \omega t^2 v_{fy} \sin \lambda \qquad (4)$$

$$y(t) = y_f + v_{fy}t - \omega t^2 v_{fx} \cos \lambda \qquad (5)$$

where $x_f$ and $y_f$ denote the horizontal point where the particle enters the 100-mbar level and $v_{fx}$ and $v_{fy}$ their horizontal reentry velocity. A scale analysis of the friction of the sliding particles with the atmosphere result in sliding times of 300-500 s, consistent with those calculated for the SL9 ejecta (Pankine & Ingersoll 1999). Reentry angles $\theta < 73°$ are discarded because (i) they would require too much time for the horizontal spread to reach the outer limits of the ejecta pattern, and (ii) the Coriolis deflection to the left is too high. Shallow impacts with $\theta > 75°$ fall back too early, with no time to deflect the horizontal components of motions by the Coriolis force during the free-falling stage. The modeled crescent structure is best fitted for a ballistic trajectory of the particulates with ejecting velocity of 7.6 ± 0.5 km s$^{-1}$, elevation angle $\theta = 70° ± 5$ ° (relative to the vertical), time aloft of 195 s (horizontal range 1,400 km) plus a sliding time of 400-500 s.

### 3. Object Trajectory and Orbit

The size of the streak's minor axis is comparable to those of Class 2a-2b SL9 impacts, but elongated in the zonal direction by a factor of 2 (e.g.- Fragment E was Class 2a, H, Q1 and R Class 2b, Fig. 2d-f). This could be due to a higher impact elevation angle, $\theta$. Assuming that the zonal length of the streak was proportional to the size of the entering body and to $sec \, \theta$ (Mac Low 1996; Zahnle 1996), a comparison with SL9 impacts where $\theta \sim 45°$, gives an elevation angle $\theta \sim 69°$, consistent with the above crescent orientation calculations. The shallower incidence angle relative to the "horizon" indicates that the body suffered initially higher ablation per unit descent altitude, and thus might have a smaller penetration level than the SL9 impacts. Assuming the impactor was an icy body entering at Jupiter's ~60 kms$^{-1}$ escape velocity, theoretical impact models (Crawford 1996; Mac Low 1996; Zahnle 1996; Korycansky et al., 2006) of SL9 fragments with similar debris structure and albedo, imply a ~0.5-km diameter. However, if the atmospheric ablation of the initial body size depends on the elevation angle as ~ $sec\ \theta$ (Crawford 1996), the pre-entry body could have been as large as ~1 km.

We ran backward numerical integrations of the orbital motion of the impacting body to constrain its nature and origin following the same procedure as Chodas & Yeomans (1996). A Monte-Carlo analysis of more than 112,000 runs was performed, starting the integrations from an impact time window of 9 to 11 UT (in steps of 2 minutes) on 2009 July 19, with pre-impact velocities ranging from 54.52 to 55.1 kms$^{-1}$ (in steps of 0.001 kms$^{-1}$) relative to Jupiter's inertial reference frame. Just as for SL9, the heliocentric orbits of the candidate impactors fell into two groups: one inside and one outside of Jupiter's orbit (semimajor axis of 5.20 AU, eccentricity of 0.048, marked with a diamond in Fig. 3). The integrations stopped in 1850 when motions become chaotic (Chodas & Yeomans 1996). The probability is 47% probability that this object impacted Jupiter directly from its heliocentric orbit (cases with impacts in the last 4 years), versus 53% that it was captured in Jovicentric orbit before impact, most probably after 1989. This differs from SL9, which was definitively captured before impact (Chodas & Yeomans 1996). To classify the orbit, we computed the invariant Tisserand parameter with respect to Jupiter for these runs (Fig. 4). Values less than 3 indicate cometary-type orbits, and values greater than 3 indicate asteroidal-type orbits. Our analysis indicates that the chance is more or less equal for the origin of this object to be in the main belt (Hilda asteroids or quasi-Hilda comet population) or from the Jupiter family comet (JFC) population. We note that the SL9 pre-capture orbit was most probably of asteroidal-type orbit (Chodas & Yeomans 1996), belonging to the quasi-Hilda family of comets.

## 4. Impact Rates at Jupiter and Future Detections

The impact rate of 0.5-1 km size bodies with Jupiter has been estimated to be 1 per 50-350 years (Fig. 5), based on a possible impact observed by Cassini in 1690 (Schenk &

Zahnle 2007), the SL9 impacts in 1994 (Hammel et al. 1995; Harrington et al. 2004), the impact crater records on the Galilean satellites (Zahnle et al. 2003; Schenk et al. 2004) and from theoretical calculations (Nakamura & Yoshikawa 1995; Kary & Dones 1996; Roulston & Ahrens 1997; Levison et al. 2000),.

The 2009 event effectively doubles the available statistical sample of well-documented collisions with Jupiter. On the sole basis of SL9 and this impact, the collision rate with Jupiter for 0.5-1 km objects is 1 per 15 years. However, accounting for the ~4-month period of bad or impossible Jupiter visibility around solar conjunction, and the typical ~2-3 month survival time of the scars for their identification in the visible (depending on the impact intensity, latitude and atmospheric wind shears), the rate could be reduced to 1 impact per decade, 5-10 times the most recent impact rate calculations as shown in Figure 5.

To test this, we calculated the detection probability of the debris left by an impact with a size >0.5 km based on the available data base with observations of the planet at visible wavelengths between 1996 and 2009 (IOPW, Hubble Space Telescope 1996-2009, Cassini flyby in 2000 and New Horizons flyby in 2007). The detection probability is assumed to be unity for all high resolution imaging for a month before the observing dates, which is a characteristic time for the reconnaissance of the debris left by a 0.5 km object. For the other cases the detection probability is assumed to follow a non-normalized Gaussian distribution centered in each apparition at Jupiter's opposition and with null probability values at Jupiter conjunctions. The full width half maximum of the Gaussian is assumed to follow the distribution of IOPW image contributions, amounting to more than 7,000 images from 2000/2001 to 2009. For campaigns before 2001 the FWHM is assumed to be 60 days as in the following years. The maximum detection probability for IOPW data is 0.35 (35% before 2001) and 0.50 (50% after 2001) accounting for the increasing number of quality observations. We find that the integrated probability of having detected an event like this from 1996 to 2009 is 40 ± 6%, equivalent to an effective impact observing time of 5.6 ± 0.8 years. The errors are calculated by increasing the IOPW effective probability up to 100% and decreasing the Cassini and New Horizons observing windows to include only the highest-resolution images. It should be noted that this is an upper limit, since high-resolution images are not likely to detect a small impact, especially at near-polar latitudes.

In addition, the temporal variation of Jupiter's declination, which makes the relevant set of amateur observers shift from the more populous northern terrestrial hemisphere to the southern one, is not taken into account.

Additionally, we performed a Monte-Carlo exploration of the probability of having an impact of a body of size larger than 500 m in Jupiter in the last 15 years based on the impact rates appearing in Figure 5 (Levison et al., 2000; Schenk and Zahnle, 2007). We find a value of 8-32%, which transforms to a 3-13% probability of observing such an impact when taking into account the effective observing time of Jupiter in the last 15 years.

Determining the statistics and probability of impacts of large bodies with Jupiter requires a continuous imaging survey. In the CCD imaging range (continuum wavelengths from 350 nm to 1 μm), the impact debris is darker than Jovian clouds, and could be identified to a size as small as ~300 km. The current large number of amateurs using CCD webcam imaging and stacking processing methods allows for a survey in much greater depth (in time and resolution on the planet) than 10 years ago, when less efficient single CCD imaging was employed, or 20 years ago when photography and visual drawing was performed by a smaller number of amateurs (Rogers 1995). This was probably why previous events were not detected. The discovery and identification of unpredicted impacts, such as the current one, could be best performed in the near-infrared methane absorption bands at 890 nm for optical CCDs and even better in near-infrared methane-hydrogen absorptions with the K band (2.12-2.3 μm), where the high-altitude aerosols make the impact features much brighter than Jupiter's primary clouds. Optimal results would be obtained by dedicated telescopes, imaging Jupiter regularly in these wavelengths, complemented by deep imaging surveys near Jupiter searching for impact bodies to allow planning and preparation for observing impacts itself, as occurred with SL9.

This work supported by the Spanish MEC AYA2006-07735 and MICIIN AYA2009-10701 with FEDER and Grupos Gobierno Vasco IT-464-07. GO and PYF acknowledge support from NASA grants to JPL. LNF was supported by NASA Postdoctoral Program at the Jet Propulsion Laboratory (Caltech, USA). JLO acknowledges AYA2008-06202-C03-01.

**FIGURES**

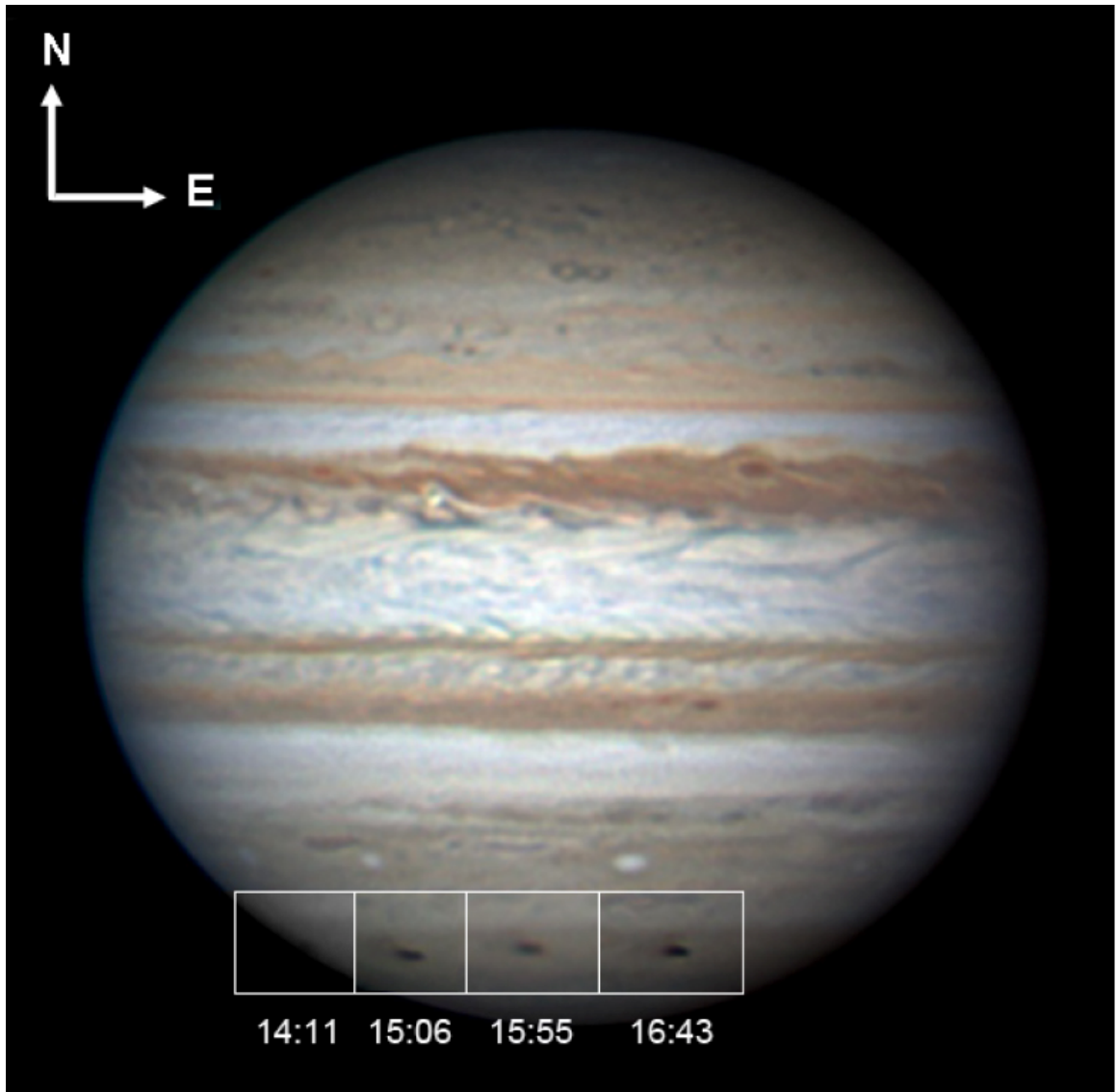

**Figure 1**
Discovery series of the impact debris obtained on 2009 July 19 at the indicated times (Newtonian telescope with a 368 mm diameter and a camera with a RGB filter covering the spectral ranges 400-700 nm). Ninox software was used for cropping and presorting of the individual frames (Wesley, 2009), with RegiStax software used for alignment and stacking (RegiStax 5, 2009).

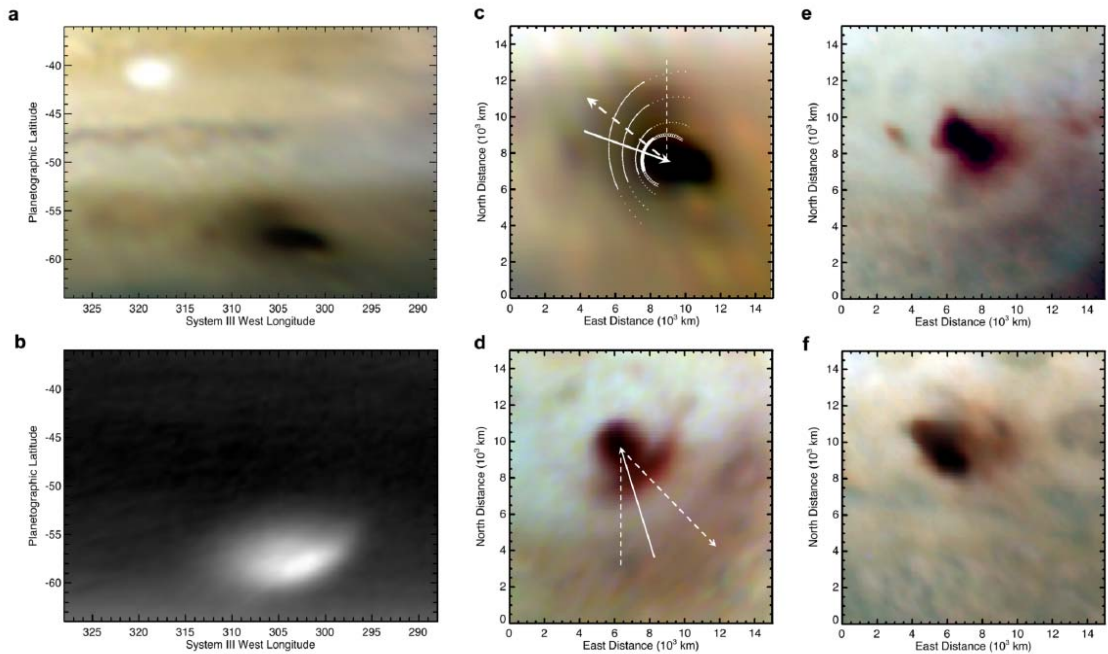

**Figure 2**

Map projections of Jupiter impact debris and comparison with SL9. Cylindrical maps: **(a)** Visible wavelengths (July 19, 16:43 UT) processed with the reconstruction algorithm PIXON (Puetter & Yahil 1999), **(b)** Near-infrared at 2.16 μm in strong methane and molecular hydrogen absorption (July 20, 11:09 UT). The feature appears smeared northeast to southwest because of the seeing conditions. Orthographic projections **(c, d, e, f)**: **(c)** 2009 July 19 impact site (as in **a**); for comparison: **(d)** the SL9 fragment E 2 hr after the impact. The continuous white arrows indicate the direction of the bolide entry, and the dashed arrows indicate the axis of symmetry of the plume ejecta. The arc curves are from the ballistic model of the ejecta with the thick arcs marking the horizontal range limits for times of 100 s, 300 s and 500 s. To assess the impact time, compare frames **(e)** and **(f)** for two similar SL9 cases that correspond to impacts R after 4 hr and Q1 after 13 hr, respectively.

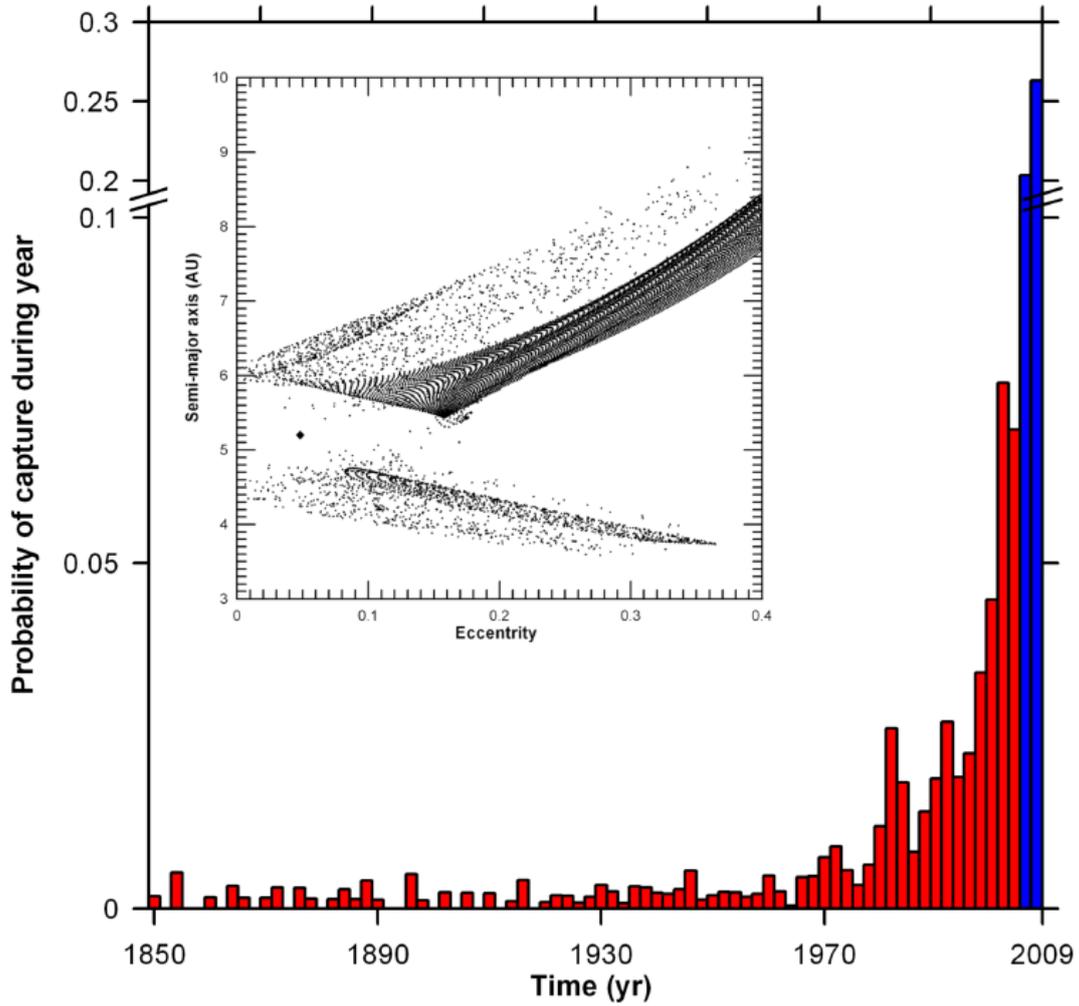

**Figure 3**
Histogram showing the probability that the July 2009 impact occurred directly from the object heliocentric orbit or was captured in any given year since 1850. The inset shows the scatter plot of possible heliocentric orbits (semi-major axis vs eccentricity) for the impacting object computed from a backward integration of the derived trajectory.

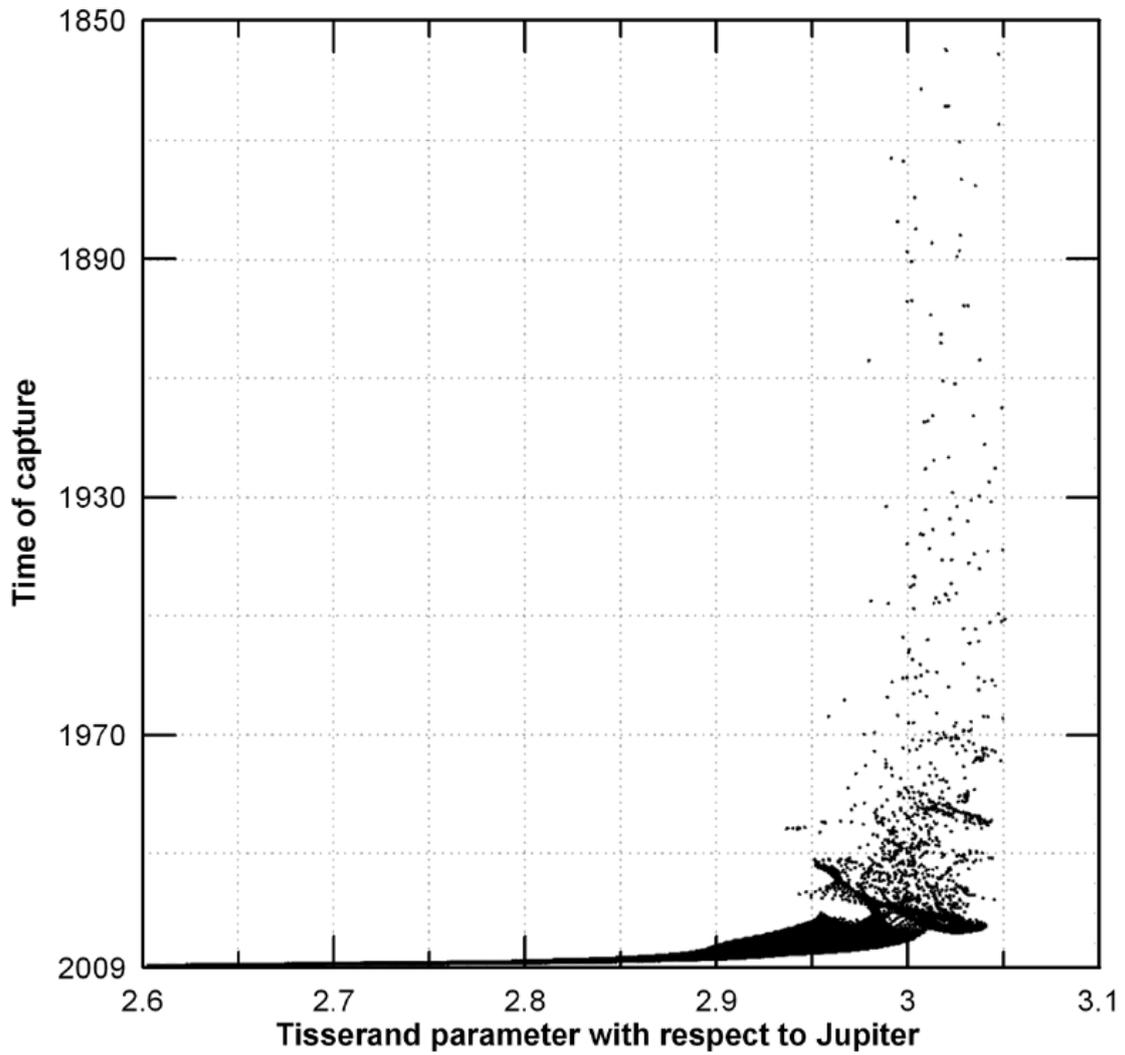

**Figure 4**
Scatter plot of the time needed for the impact object to reach a distance of 2 Astronomical Units from Jupiter vs. the Tisserand parameter with respect to Jupiter. Beyond 2 A.U.it is assumed that the orbital elements of the body are not significantly modified by Jupiter.

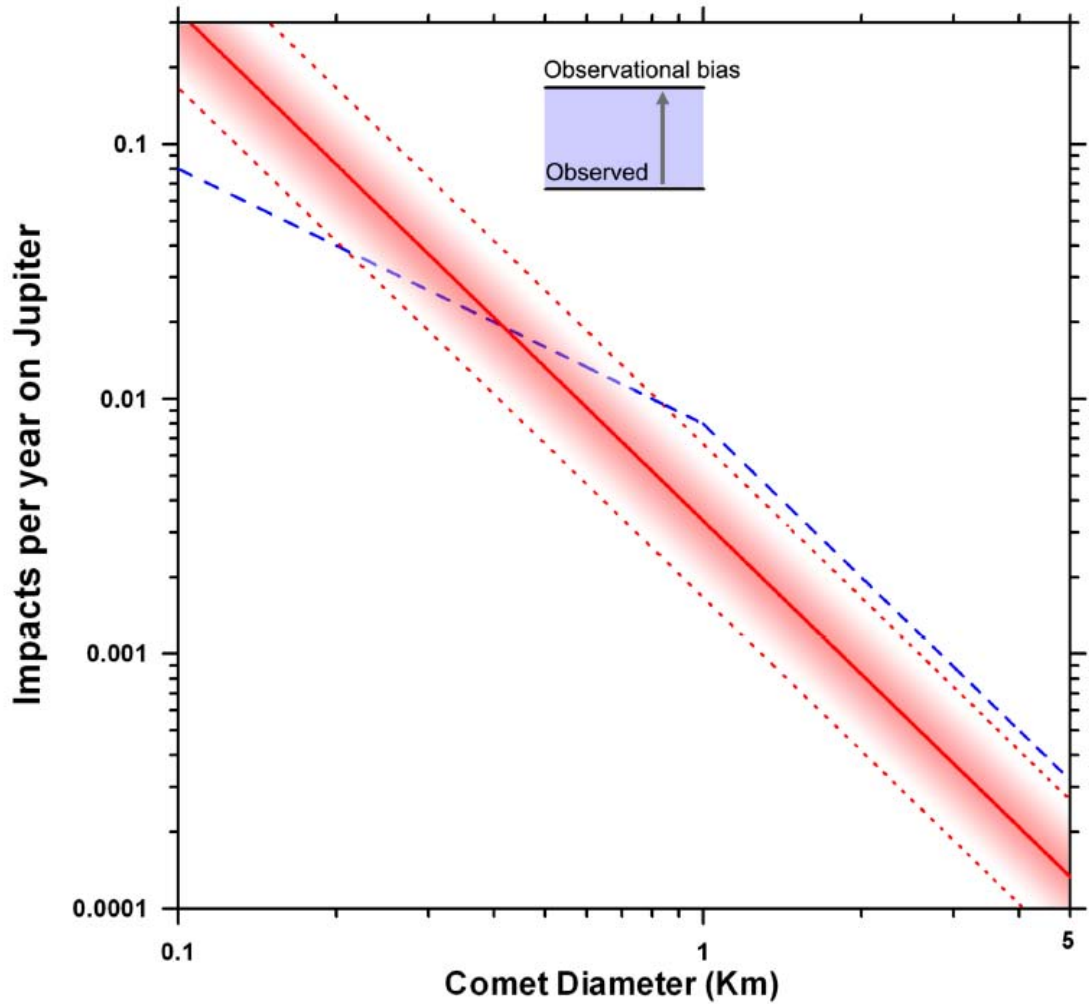

**Figure 5**
Cumulative impact rates per year at Jupiter as a function of the impacting object size compared to the two most recent impacts (SL9 and 2009, upper blue box). The blue dashed line is obtained with data taken from Schenck et al. (2004). The red continuous line corresponds to the scenario presented by Levison et al. (2000). The uncertainty is represented by the red dotted line boundaries obtained by multiplying the mean impact rates by 2 and 0.5.